\documentclass[12pt, aps,pra,preprint,showpacs]{revtex4-1}
\usepackage{graphicx,color,subfigure}
\newcommand\bom{{\mbox{\boldmath $\omega$}}}

\begin{document}
\title{Helicity and topology of a small region of quantum vorticity}
\author{
M. Mesgarnezhad$^1$, R.G. Cooper$^1$, A.W. Baggaley$^1$ and C.F. Barenghi$^1$}
\address{$^1$ School of Mathematics and Statistics and
Joint Quantum Centre Durham-Newcastle,
Newcastle University, Newcastle upon Tyne NE1 7RU, UK}
\email{m.mesgarnezhad1@newcastle.ac.uk}

\begin{abstract}
We numerically study the evolution of a small turbulent
region of quantised vorticity
in superfluid helium, a regime which can be realised
in the laboratory. We show that the turbulence achieves a fluctuating
steady-state in terms of dynamics (energy), geometry (length, writhing)
and topology (linking). 
We show that, at any instant, the turbulence consists of many
unknots and few large loops of great geometrical and topological
complexity.
\end{abstract}
\maketitle

\vspace{2pc}


%\maketitle
 

\newpage

\section{Introduction}

Tangled filamentary structures occur in many physical
systems, from ropes to optics \cite{Padgett2011} to DNA \cite{Vologodskii1998}. 
Concentrated field lines in 
fluids \cite{Kida1994} and plasmas \cite{Priest2000} 
(e.g. vortex lines and magnetic field lines)
are another important example.
Such lines undergo reconnection events, which are associated
with energy losses. In the limit of no dissipation, the governing equations
of motion (the Euler equation and the magnetic induction equation
in the frozen field approximation respectively) 
preserve the topology of the lines.
In this limit, helicity and magnetic helicity are conserved quantities.
Recent work suggests that, in the case of small dissipation, helicity
is partially preserved \cite{Kleckner2015}. The aim of this work is to explore
this partial preservation of helicity in a context
where vortex lines are not mathematical abstractions but have a real 
physical meaning:  this is the context of quantum fluids, notably
superfluid 
liquid helium ($^4$He and $^3$He) and atomic Bose-Einstein condensates.
The numerical experiment which we describe explores
the relation between geometry, dynamics and topology of a small region
of turbulence in superfluid helium.

\section{Quantum vorticity}

Quantum fluids are characterised by zero viscosity
and the quantisation of the circulation. 
The first property makes 
superfluids similar to inviscid Euler fluids of 
traditional textbooks. The
second property arises from the existence of a complex macroscopic
wavefunction 
$\Psi({\bf x},t)=\sqrt{ n({\bf x},t)} e^{i\phi({\bf x},t)}$
where ${\bf x}$ is the position, $t$ is the time,
$n({\bf x},t)$ is the number density;  according to the
Madelung transformation \cite{primer}, the superfluid velocity is proportional
to the gradient of the phase:
% $\phi({\bf x},t)$ \blue{of $\Psi({\bf x},t)$}:

\begin{equation}
{\bf v}({\bf x},t)=\frac{\hbar}{m}\nabla \phi,
\end{equation}

\noindent
where $m$ is the mass of the relevant boson 
($m=6.6 \times 10^{-27}~\rm kg$ for $^4$He), $\hbar=h/(2 \pi)$ 
and $h=6.6 \times 10^{-34}~\rm J~s$ is Planck's constant.
The single-valuedness of $\Psi$ implies that
the circulation of the superfluid
velocity field around a closed path $C$ is either
zero or a multiple of the quantum of circulation $\kappa=h/m$:

\begin{equation}
\oint_C {\bf v} \cdot {\bf dr}=n \kappa,
\label{eq:kappa}
\end{equation}

\noindent
($n=0, \pm 1, \pm 2, \cdots$).
Nonzero circulation occurs when the path $C$ encloses 
a vortex line; in this case $\Psi=0$ on the axis of the vortex line. 
Since multi-charged {($\vert n \vert >1$) vortices are
unstable, we are concerned only with the case $n=\pm 1$.
Around the axis of the vortex line there is a thin tubular region of 
depleted density of radius $a_0 \approx 10^{-10}~\rm m$ in $^4$He; 
in this region, the density $n({\bf x},t)$ drops from its 
bulk value at infinity to zero on the vortex axis.
In summary, a vortex line is a hole with superfluid circulation 
around it.
As in classical fluid dynamics, quantum vortex lines are either closed loops 
or terminate at boundaries.

An isolated vortex is a stable topological defect which
does not decay, unlike vortices in ordinary viscous fluids. Numerical
simulations of the governing Gross-Pitaevskii equation for $\Psi$
show that vortex lines reconnect when they come close to each
other \cite{Koplik1993,Zuccher2012,Allen2014}, as observed in experiments 
in superfluid helium \cite{Bewley2008} and, more recently,
in atomic condensates \cite{Serafini2015}.
In this process, some of the kinetic energy 
of the vortices
is turned into density waves \cite{Leadbeater2001,Zuccher2012}.
Evidently, vortex reconnections are of particular important for the 
dynamics of turbulence in superfluid helium \cite{PNAS2014}
and in atomic Bose-Einstein condensates \cite{Tsatsos2016}.

In typical experiments the average distance between vortices,
$\ell \approx 10^{-4}$ to $10^{-2}~\rm m$, is many orders of magnitude larger than the vortex core radius $a_0$. It is therefore appropriate
to model quantum vortices as
(closed) space curves ${\bf s}(\xi,t)$ (where $\xi$ is arc-length)
of infinitesimal thickness. In a pure superfluid (e.g. liquid helium
at temperatures below $1~\rm K$) thermal excitations are {particularly}
negligible,
and the vortex moves according to \cite{Saffman1992}

\begin{equation}
\frac{d{\bf s}}{dt}={\bf v}_{self}({\bf s}),
\label{eq:Helmoltz}
\end{equation}

\noindent
where the self-induced velocity
${\bf v}_{self}$ is given by the classical Biot-Savart law:

\begin{equation}
{\bf v}_{self}({\bf s})=
-\frac{\kappa}{4 \pi} \oint_{\cal L} 
\frac{({\bf s} -{\bf r}) \times {\bf dr}}
{\vert {\bf s} - {\bf r} \vert^3}.
\label{eq:BS}
\end{equation}

\noindent
(the line integral extending over the entire vortex configuration
$\cal L$).
Numerical simulations of quantum vortex lines are based on Lagrangian
discretisation of the lines \cite{Schwarz1988}. 
The number of discretisation points along
a line varies, as more/less points are required in regions of
high/low curvature.
The Biot-Savart
integral is de-singularised in a standard way \cite{Schwarz1988}
(based on the distance $a_0$), and
an algorithmic reconnection procedure is implemented \cite{Baggaley-recon}. 
The numerical
method is standard and has been published 
in the literature \cite{AFT2010,Baggaley-tree}.

In this work, we are concerned with  (experimentally easily
accessible)
high temperatures ($T> 1~\rm K$). In this regime, the thermal excitations form
a viscous fluid (called the normal fluid) of velocity field ${\bf v}_n$
which exchanges energy with the
vortex line via a mutual friction force \cite{BDV1983}. 
Eq.~(\ref{eq:Helmoltz})
requires modifications, and becomes \cite{Schwarz1988}

\begin{equation}
\frac{d{\bf s}}{dt}={\bf v}_{self} 
+\alpha {\bf s}' \times ( {\bf v}_n-{\bf v}_{self} )
-\alpha' {\bf s}' \times [{\bf s}' \times ( {\bf v}_n-{\bf v}_{self}].
\label{eq:Schwarz}
\end{equation}

\noindent
Here 
$\alpha$ and $\alpha'$ are small temperature-dependent friction
coefficients arising from the interaction of the vortex lines
with the thermal excitations which make up the normal fluid 
(if $T \to 0$ then $\alpha \to 0$ and 
$\alpha' \to 0$, recovering Eq.~(\ref{eq:Helmoltz})).
It must be stressed that the friction force is a two-way route 
to exchange energy between the normal fluid and the superfluid 
vortex lines. 
If a section of a vortex line is exposed to normal flow ${\bf v}_n$
locally aligned in the same direction of the superfluid vorticity
and with intensity larger than a certain
critical value, infinitesimal perturbations on that section 
of the vortex line will grow in length
in the form of Kelvin waves (helical displacements of the vortex axis
away from its initial position); this effect is called the
Glaberson-Donnelly instability \cite{Glaberson}. Vice-versa, if ${\bf v}_n$
is perpendicular to the vortex line or if ${\bf v}_n=0$,
the friction will dissipate the Kelvin waves.
Similarly, if ${\bf v}_n$ blows along the direction of propagation of
a superfluid vortex ring, the ring will increase its radius; vice-versa,
if ${\bf v}_n$ blows in the opposite direction or if ${\bf v}_n=0$, the ring
will shrink and vanish.

\section{Turbulent quantum vorticity}

Our aim is to numerically simulate a state of turbulence of vortex lines
which has two properties: 
(i) it is in a statistical steady-state (independent of the arbitrary
initial condition, unlike previous work \cite{BRS2001} which was concerned 
with comparing energy and complexity during an initial transient), 
and (ii) is away from (hard or periodic)
boundaries (so that vortex lines
are closed loops and the definition of linking
is simple and unambiguous).

It is not trivial to satisfy (i) and (ii) at the same time.
Most experiments
and numerical simulations of quantum turbulence have been performed
within hard or periodic boundaries. In these studies,
after an initial transient, 
a steady-state regime is achieved and the vortex lines
fill the entire domain, which means that some vortex lines are not closed
loops but terminate at
the boundaries. 
Because of the Lagrangian discretisation,
calculations can also be performed in open, infinite domains, 
but in this case the vortex length will grow without
limit. 

The calculations which we present here model 
experiments \cite{Smith1981,Milliken1982} 
in which ultra-sound waves were focused at the centre
of the experimental cell creating vortex lines away from boundaries.
In this experiments, vortex lines which left the central region
decayed due to friction with the normal fluid which was stationary
far from the central region. To model this experimental configuration
we impose a time-dependent, space dependent velocity field ${\bf v}_n$
in the form of random waves \cite{Baggaley2011-stats} near the
centre of the computational domain; away from the centre, we impose
that the normal fluid velocity decays exponentially - see Fig.~(\ref{fig1}). 
In the central region we place 
typically 40 vortex rings (planar unknots) 
as seeding initial condition at $t=0$. 
The images which we present mainly refer to
a numerical simulation in which the initial rings have the same radius
but random orientation, and their centres are shifted
according to a normal distribution, but it is important to note that
we have performed simulations starting from different initial conditions
(e.g. rings of random positions and sizes) and longer times in statistical
steady-state regime, and the results appear to be 
independent of the precise initial vortex configuration. 
All numerical simulations are
performed at a temperature $T=1.9~\rm K$ (corresponding to
$\alpha= 0.206$ and $\alpha'= 8.34 \times 10^{-3}$) which is
typical of experiments.

During the evolution, we observe that
the vortex length rapidly increases in the central 
region due to the Donnelly-Glaberson instability,
resulting in Kelvin waves and vortex reconnections 
(which continuously change the number of vortex loops);  
a turbulent tangle of vortex lines quickly grows, 
as shown in Fig.~(\ref{fig2}).

We have not measured directly
the distribution of angles $\theta$ between reconnecting
vortex lines during our numerical simulations, but recall the results 
of a previous study\cite{Sherwin} which determined $\theta$ for two
different fully-developed turbulent regimes.
In the first regime, the turbulence exhibited a classical Kolmogorov cascade
(energy spectrum concentrated at the largest length scales followed by
$k^{-5/3}$ scaling at larger wavenumbers $k$), and
the distribution of the reconnecting angles $\theta$ peaked at small
angles ($\theta \approx \pi/8$ in Fig.~9 of cited reference). Indeed, the
vortex tangle visibly contained metastable bundles of parallel
vortices, making parallel reconnections relatively more frequent.
In the second regime, the turbulence did
not exhibit any sign of a classical cascade (energy spectrum concentrated
at the mesoscales and $k^{-1}$ scaling at larger $k$), the distribution of
reconnection angles peaked at larger angles closer to 180 degrees
($\theta \approx 7 \pi/8$ in Fig.~9 of cited reference). Indeed,
the vortex tangle lacked bundles and looked more random, making
antiparallel reconnections relatively more frequent.
We believe that the relatively small vortex configurations
which we study here are more similar to the second regime: they are
visibly more random-looking and lack the large separation of length scales
between the average intervortex spacing and the system's size for the
classical Kolmogorov spectrum to develop. We expect therefore
that antiparallel reconnections are relatively more frequent.

After a quick initial transient, a balance is reached between 
vortex generation and vortex decay in the central region, and the 
vortex length saturates.  Vortex loops
which drift too far away from the central region decay
due to friction with the stationary normal fluid in the outer region;
this effect
helps create the desired saturated, localised region of quantum turbulence.
characterised by fluctuations of the vortex length $\Lambda$ about an average
value, as shown in Fig.~(\ref{fig3}).
The small loops which drift away and escape from the
central region are either rings or slightly deformed
rings (that is to say unknots). Larger more knotted structures are slower
and cannot escape easily from the central region (if they did,
friction with the normal fluid would lead them to shrink, thus eventually
reconnect, turning them into unknots.

In Fig.~(\ref{fig2}), note that the tangle does not necessarily remain 
isotropic during the evolution: given the random nature of the driving ${\bf v}_n$
and the exponential growth of the Donnelly-Glaberson instability,
a particular realisation of the synthetic turbulence may generate a vortex configuration
which moves off-centre or does not extend by the same amount 
in all directions.

\section{Geometry and topology of the turbulence}

In this section we analyse the dynamical, geometrical
and topological properties of
the steady-state regime of quantum turbulence which we have achieved. 
We have checked that not only the length, but also the kinetic
energy $E$ of the vortex lines saturates to an average value, see
Fig.~(\ref{fig4}). The energy is evaluated from \cite{Saffman1992}

\begin{equation}
E=\int_V {\bf v} \cdot {\bf r} \times \bom dV,
\label{eq:E1}
\end{equation}

\noindent
(where $V$ is volume). In our case, assuming
that ${\bf v} \to 0$ at infinity, since
vorticity is concentrated along filaments, Eq.~(\ref{eq:E1}) reduces to

\begin{equation}
E=\kappa \oint_{\cal L} {\bf v} \cdot {\bf r} \times {\bf ds}',
\label{eq:E2}
\end{equation}

\noindent
where the line integral extends over the entire vortex tangle $\cal L$. 
A quantity which is often reported in the literature as a 
measure of the turbulent intensity is the vortex
line density (length of vortex line per unit volume), defined as $L=\Lambda/V$.
Without boundaries, the volume which contains the lines is not a well-defined
quantity; we estimate it as the sphere which contains $95 \%$ of the vortex
lines. In the simulation shown in Fig.~(\ref{fig3}), we obtain
$L \approx 2500~\rm cm^{-2}$

Further analysis is performed using the concept of {\it crossing numbers}.
At a given instant, the turbulent vortex tangle consists of a number of 
vortex loops: ${\cal L}= {\cup}_j {\cal L}_j$. 
We project the tangle $\cal L$ onto a given 2D plane. The projected
tangle is a self-intersecting curve: the points of intersection correspond
to apparent crossings of $\cal L$ as seen from the line of sight of the
projection. Since each loop is oriented (by the sense of rotation of the 
superfluid velocity), we can assign values $\epsilon_k=\pm 1$ to each
point of intersection $k$ according to standard convention, see
Fig.~(\ref{fig5}). The total number of crossings, $k$, is plotted vs time in Fig.~(\ref{fig6}); this is the simplest measure of the complexity of the tangle. We can also readily define the {\it writhing number}
of the entire tangle as

\begin{equation}
W=\left \langle \sum_{k \in {\cal L}} \epsilon_k \right \rangle,
\label{eq:W}
\end{equation}

\noindent
where the symbol $\langle  \cdots \rangle$ denotes the average over a 
number of projections to make the result independent of a particular
projection \cite{BRS2001}. Ideally, the writhing number should be
estimated by integrating over all solid angles; in practice, this would be
computationally expensive. Numerical experiments \cite{BRS2001} 
suggest it suffices to
average over a small number of projections. The results which we report
are calculated simply} by projecting over 
the three Cartesian planes.
We find - see Fig.~(\ref{fig7}) - that the writhing number of the tangle
achieves a statistically steady-state too.

Other geometrical and topological properties of
individual vortex loops ${\cal L}_j$ or of the entire tangle $\cal L$
can be evaluated by suitably conditioning Eq.~(\ref{eq:W}).
For example, the writhing number $W_j$ of an individual
loop ${\cal L}_j$ is 
obtained by summing the crossing numbers restricted to vortex strands
which belong to that particular loop:

\begin{equation}
Wr_j=\left \langle \sum_{k \in {\cal L}_j} \epsilon_k \right \rangle.
\label{eq:Wj}
\end{equation}

Another quantity of interest is the {\it linking number} 
$Lk_{i,j}$ between loops
${\cal L}_i$ and ${\cal L}_j$, which is defined as

\begin{equation}
Lk_{i,j}= \frac{1}{2} \sum_{k \in {\cal L}_i \cap {\cal L}_j} \epsilon_k.
\label{eq:Li}
\end{equation}

\noindent
This quantity, computed from a single arbitrary projection, is the
same for any projection, as it can be easily verified \cite{Kauffman1987}. 
A measure of the topological complexity of the turbulence is the
the {\it total linkage} $Lk$ of the tangle, defined as

\begin{equation}
Lk= \sum_i \sum_{j\neq i} \vert Lk_{i,j} \vert.
\label{eq:Ltot}
\end{equation}

\noindent
The physical importance of the linkage is that only vortex reconnections
can undo the linking between two loops and each
reconnection has an energy cost (in terms of sound waves emitted).
The total linkage vs time is shown in Fig.~(\ref{fig8}). It is
apparent that the fluctuations of the linkage 
are relatively large, as few reconnections make a relatively large
change for a small vortex configuration like ours, but the turbulence
settles to an average value of linkage, which is 
${\bar{Lk}}= 54.07$ in Fig.~(\ref{fig8}). The spontaneous formation
of links has also been observed in a numerical simulation of decaying
turbulence using the Gross-Pitaevskii equation \cite{Villois2016}.

Now we turn the attention to the {\it helicity}. In a classical
fluid, the helicity is defined as \cite{Moffatt1969}

\begin{equation}
H=\int_V {\bf v} \cdot {\bom} dV,
\label{eq:H}
\end{equation}

\noindent
where $\bom$ is the vorticity. What should be the correct
definition of helicity in a quantum fluid is a hot topic 
in the current literature \cite{Brachet2016,Zuccher2015,Hanninen2016}. 
The difficulty is that,
on one hand, we would like a definition of helicity which is 
consistent with the classical definition of helicity, and,
on the other hand, in a quantum fluid 
vorticity $\bom$ is zero everywhere with the exception of
the axes of vortex lines; more precisely,  $\bom$ is a delta function 
on the vortex axis, which unfortunately
is also the location where the velocity is undefined.
%It must be stressed that the wavefunction $\psi$ is regular
%everywhere; what becomes singular on a vortex axis
%is the Madelung transformation, not the Gross-Pitaevskii equation. 
It is worth remarking that in a classical fluid
one thinks of the vortex core as a small bundle of mathematical
vortex lines; this allows an interpretation of helicity which contains 
internal twist.
In a quantum fluid, however, there is {\it only one vortex line}
(on the axis of the vortex - everywhere else the flow is potential).
For the sake of simplicity, we follow Scheeler et al
\cite{Scheeler2014}
and interpret $H$ as {\it centreline helicity} defined as

\begin{equation}
H= \kappa^2 \left( 
2 \sum_i \sum_{j \neq i} Lk_{i,j} + \sum_i Wr_i
\right).
\label{eqL:H2}
\end{equation}

\noindent
An important physical property of the centreline helicity is that
it is invariant under anti-parallel reconnections characteristic
of fluid flows \cite{Laing2015}. Fig.~(\ref{fig9}) shows that the
centreline helicity remains approximately constant during the
time evolution. 
After the initial transient, the average value is
${\bar H} \approx -96.39 \kappa^2$.

We also characterise the complexity of the turbulence by
computing the relative distribution of vortex loops with a
given value of writhe. The relative writhe distribution
suggests that
although most loops have a small value of $\vert Wr_j \vert$,
at any instant there is a consistent number of loops with large value
of writhe 
 (see Fig.~\ref{fig10})
in the approximate range $50 < \vert Wr_j \vert < 300$.
Closer investigation confirms that these complex loops with
more coils and twists than the average loop
keep forming, decaying and reforming. Examples of vortex loops
with small and large values of $\vert Wr_j \vert$ are shown in 
Fig.~(\ref{fig11}). 

The geometrical
complexity described by the writhe spectrum
seems associated with a similar topological complexity. A simple way to
quantify the topology of a vortex loop is to determine the
order $A$ of its Alexander polynomial \cite{Alexander1928}. 
The Alexander polynomial $\bigtriangleup (\tau)$ is a topological invariant 
which can be easily computed by labelling segments of a loop between 
under-crossings when projected into a plane, followed by assigning coefficients to the relevant entries of 
a matrix for each segment, and then finding the determinant of the matrix 
with any single row and column removed \cite{Livingstone1993}. 
Given the vortex configuration at any time $t$, we compute the Alexander
polynomial of each vortex loop and find its order $A$.
For example the Alexander polynomial of the trivial unknot 
is $\bigtriangleup (\tau)=1$ hence its order is $A=0$.
The simplest knot is the trefoil ($3_1$) knot, which has Alexander polynomial 
$\bigtriangleup (\tau)=1-\tau+\tau^2$, hence its order is $A=2$. 
Any vortex loop which has an Alexander polynomial of order $A > 0$ is knotted,
however the converse is not necessarily true: a long-standing problem of 
knot theory is the lack of a unique method of distinguishing knots from each other.
In particular, the Alexander polynomial is not unique to a particular knot type.
For example there exist knots which have the same Alexander polynomial as the 
unknot \cite{DesCloizeaux1979}, so the fact that a vortex loop has an Alexander 
polynomial of order $A=0$ does not necessarily imply that it is an unknot.
Nevertheless, the order of the Alexander
polynomial of a loop is a more instructive measure of the loop's topology than
its writhe.

The distribution of values of $A$ is similar to the distribution
of values of $\vert Wr_j \vert$: we find that, at any instant,
most vortex loops are unknots, but,  there are
always some loops with a high degree of topological complexity in the 
approximate range
$25 < A < 125$. Fig.~\ref{fig12} 
shows the distribution of values of $A$ for the three most complex knots
in a given vortex configuration (that is, at each instant $t$ for a sample
of values of $t$): the figure confirms the robustness of the finding - in any 
vortex configuration there is always a small but consistent number of loops
with nontrivial topology. 
%In comparison, for random
%poligons the probability of a knot decreases exponentially 
%with complexity \cite{%Shimamura2002}.
Some examples of vortex loops with intermediate and large values of $A$ which we
observed in our numerical simulations are shown in Fig.~(\ref{fig13}).

Finally, Fig.~(\ref{fig14}) shows the relationship between the length of
vortex loops, $\Lambda_j$, and the order of their Alexander polynomials, $A$. 
It is apparent that complex knots tend to be long knots. The inset plots the
data on linear-log scales and suggests that $A$
increases roughly exponentially with $\Lambda_j$.
It is interesting to remark
that the probability that a DNA molecule (modelled as a random polygon)
is unknotted decreases exponentially with length toward 
zero\cite{Arsuaga2002}.

\section{Conclusion}

In conclusion, we have simultaneously related geometry, dynamics
and topology of a small statistically-steady
turbulent region of quantum vorticity away from
boundaries; we have chosen such a regime because it
simplifies the calculation of topological properties
and can be generated in the laboratory.
We have shown that centreline helicity is preserved in a
statistical steady-state regime like energy, length, writhing and linking. 
By examining the
writhe and the order of the Alexander polynomial of vortex loops, we
have found that most of the loops are topologically
trivial (unknots), but there are always some vortex loops of great geometrical
and topological complexity.

Further work will study how the complexity increases with the intensity of
the turbulence and look for scaling laws. 
Another line of further work would be to replace the driving 
random waves with a Beltrami flow, seeking
to induce more helicity in the turbulence.  The saturation process would
proceed as described here, but we would have less relative fluctuations of
helicity.

\section*{Acknowledgments}
We thank Prof DW Sumners and Dr A Duncan for discussions, and EPSRC (grant EP/I019413/1) for financial support. 

%%%%%%%%%%%%%%%%%%%%%%%%%%%%%%%%%%%%%%%%%%%%%%%%%%%%%%%%%%%%%%%%%%

%Alberto Villois, Giorgio Krstulovic, Davide Proment, Hayder Salman 
%arXiv:1604.03595 (2016)
%A Vortex Filament Tracking Method for the Gross-Pitaevskii 
%Model of a Superfluid

%Alberto Villois, Davide Proment, and Giorgio Krstulovic
%Phys. Rev. E 93, 061103(R) (2016)
%Evolution of a superfluid vortex filament tangle driven by 
%the Gross-Pitaevskii equation
%%%%%%%%%%%%%%%%%%%%%%%%%%%%%%%%%%%%%%%%%%%%%%%%%%%%%%%%
\newpage
\begin{figure}%
\centering
\includegraphics[width=0.9\linewidth]{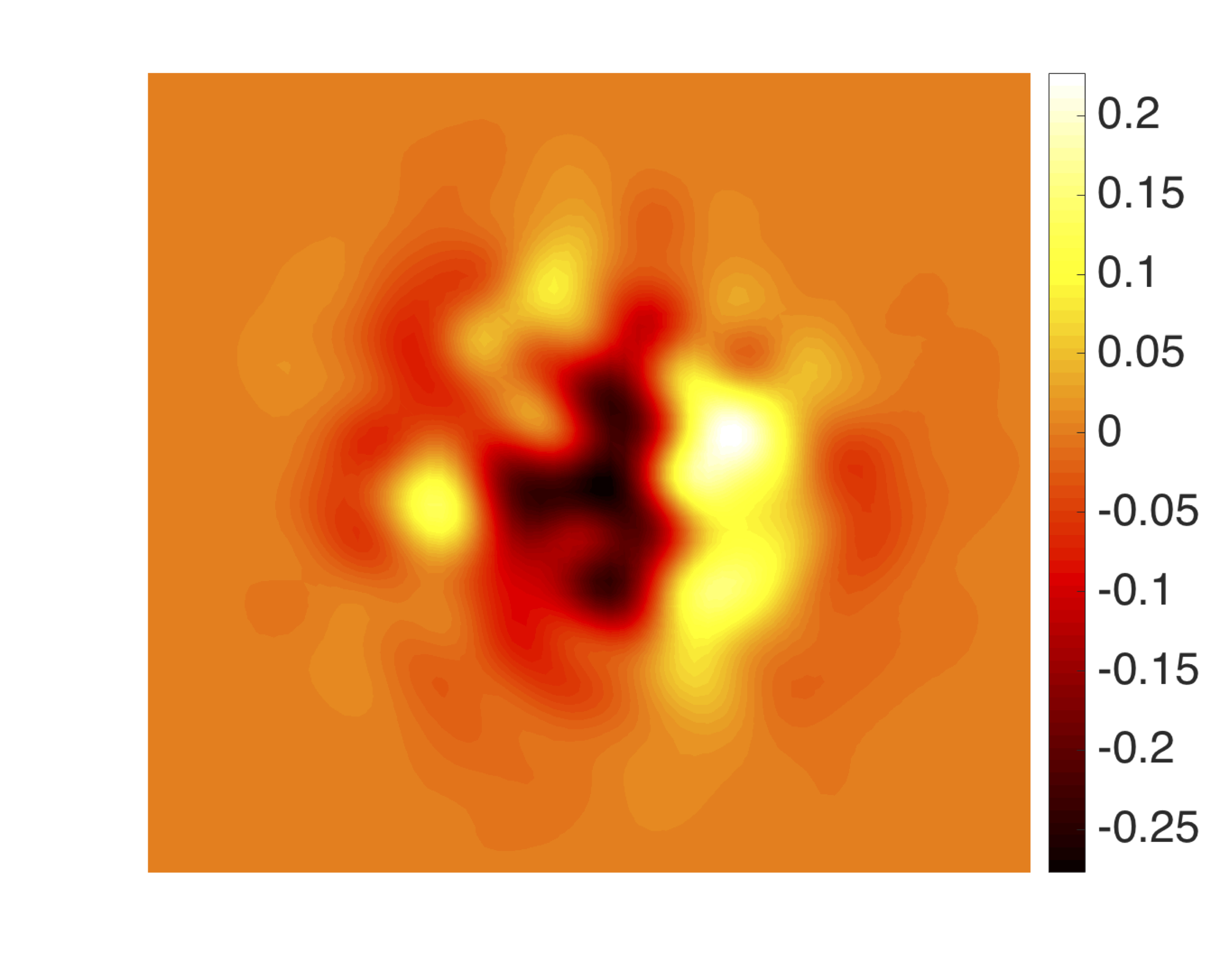}
\caption{Intensity of the normal fluid velocity ${\bf v}_n$ at
an arbitrary time plotted on the
$z=0$ plane.}
\label{fig1}
\end{figure}

\newpage
\begin{figure}%
\centering
\includegraphics[width=0.45\linewidth]{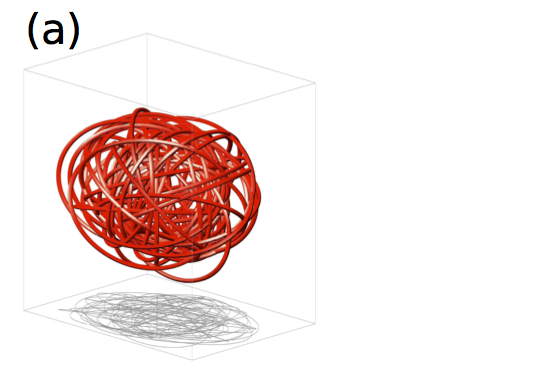}\qquad
\includegraphics[width=0.45\linewidth]{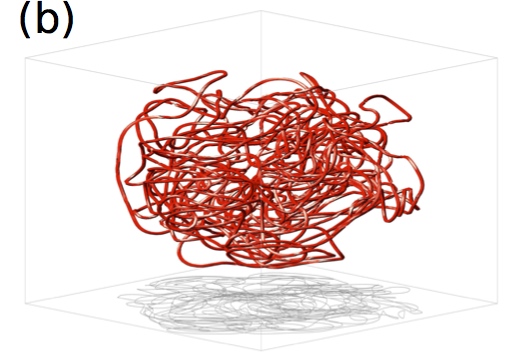}\\
\includegraphics[width=0.45\linewidth]{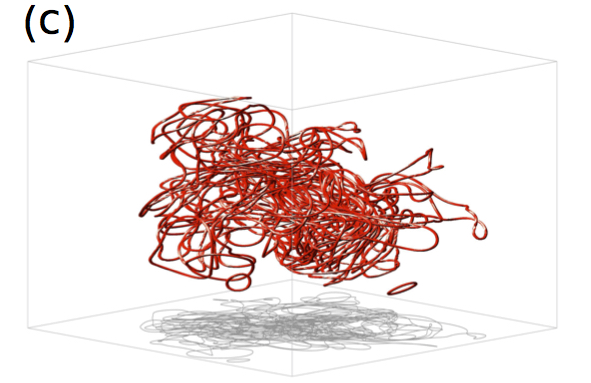}\qquad
\includegraphics[width=0.45\linewidth]{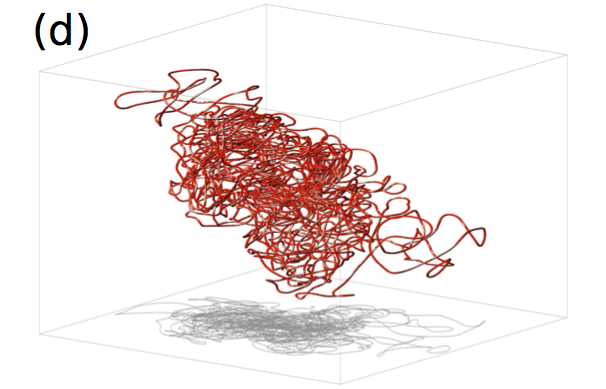}\\
\includegraphics[width=0.45\linewidth]{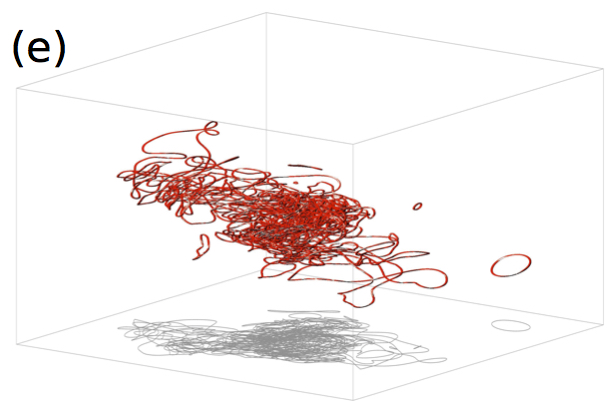}\qquad
\includegraphics[width=0.45\linewidth]{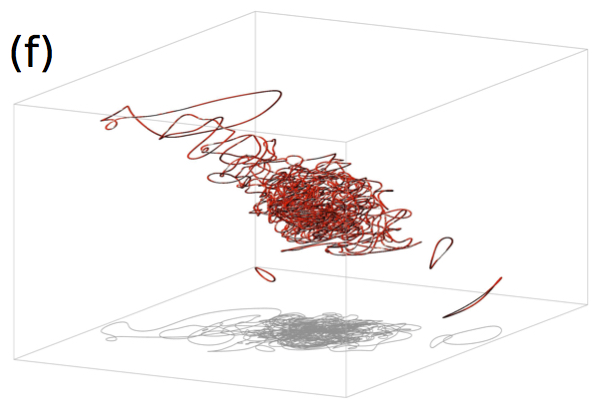}
\caption{Typical time evolution of the vortex tangle. The vortex lines are the red curves enclosed in a box with shadows for visualisation purposes only. The figure shows the vortex tangle at \newline (a): t=0.00s in the region $-0.02 \leq x \leq 0.02$, $-0.02 \leq y \leq 0.02$, $-0.03\leq z \leq 0.02$ 
\\
(b): t=0.04s in the region $-0.08 \leq x \leq 0.07$, $-0.08\leq y \leq 0.06$, $-0.09 \leq z \leq 0.08$ 
\\
(c): t=0.20s in the region $-0.11 \leq x \leq 0.10$, $-0.11 \leq y \leq 0.10$, $-0.12 \leq z \leq 0.13$ 
\\
(d): t=0.80s in the region $-0.13 \leq x \leq 0.16$, $-0.14\leq y \leq 0.15$, $-0.19 \leq z \leq 0.16$ 
\\
(e): t=1.60s in the region $-0.16 \leq x \leq 0.18$, $-0.15 \leq y \leq 0.20$, $-0.29 \leq z \leq 0.29$  
\\
(f): t=4.00s in the region $-0. 25\leq x \leq 0.23$, $-0.25\leq y \leq 0.22$, $-0.30 \leq z \leq 0.27$.  }
\label{fig2}
\end{figure}

\newpage
\begin{figure}%
\centering
\includegraphics[width=0.9\linewidth]{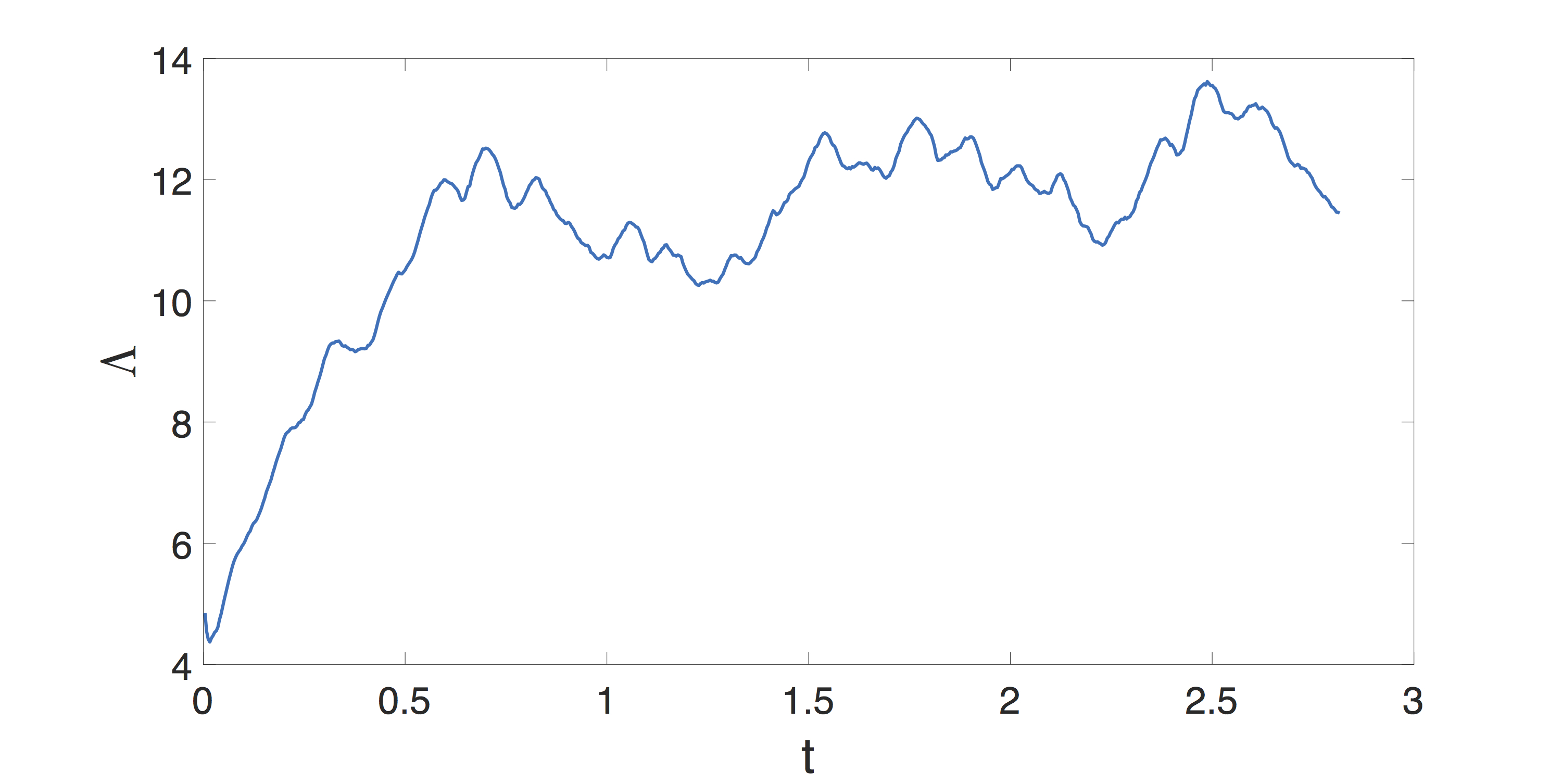}
\caption{Length $\Lambda$ of the vortex configuration (in $\rm cm$) vs
time $t$ (in $\rm s$). At $t=0$, the initial length of this realisation
is $\Lambda(0)=4.0\rm cm$ and the average length for $t>0.6$ is $\bar \Lambda = 11.77\rm cm$.
}
\label{fig3}
\end{figure}

\newpage
\begin{figure}%
\centering
\includegraphics[width=0.9\linewidth]{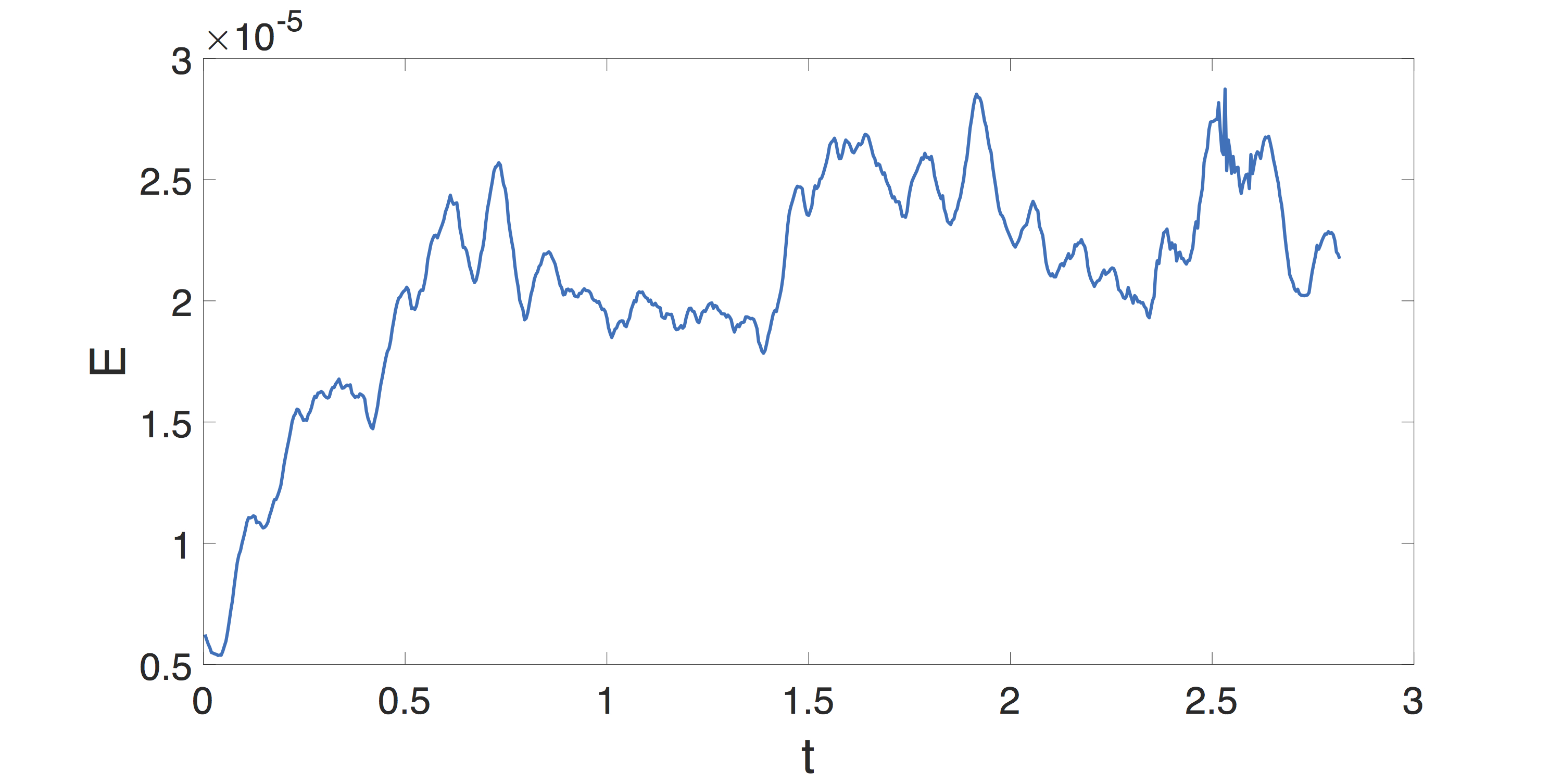}
\caption{Kinetic energy $E$ of the vortex configuration 
(arbitrary units) vs
time $t$ (in $\rm s$) corresponding to Fig.~(\ref{fig2}). 
%The average energy $\bar E = 2.1678 \times 10^5$.
The average energy for $t>0.6$ is $\bar E = 2.20 \times 10^{-5}$.
}
\label{fig4}
\end{figure}

\newpage
\begin{figure}[htp]
\centering
\includegraphics[width=0.4\linewidth]{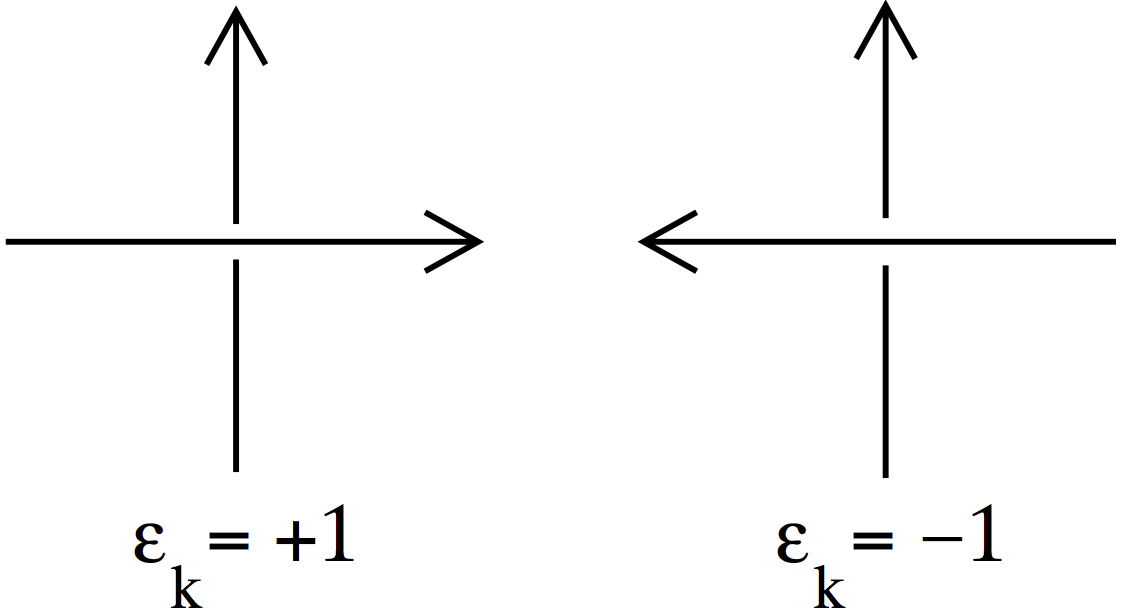}
\caption{
Each point of intersection $k$ of the projected tangle is assigned
a crossing number $\epsilon_k= \pm 1$ depending on the relative
orientation of the vortex lines as schematically shown here.
}
\label{fig5}
\end{figure}

\newpage
\begin{figure}
\centering
\includegraphics[width=0.9\linewidth]{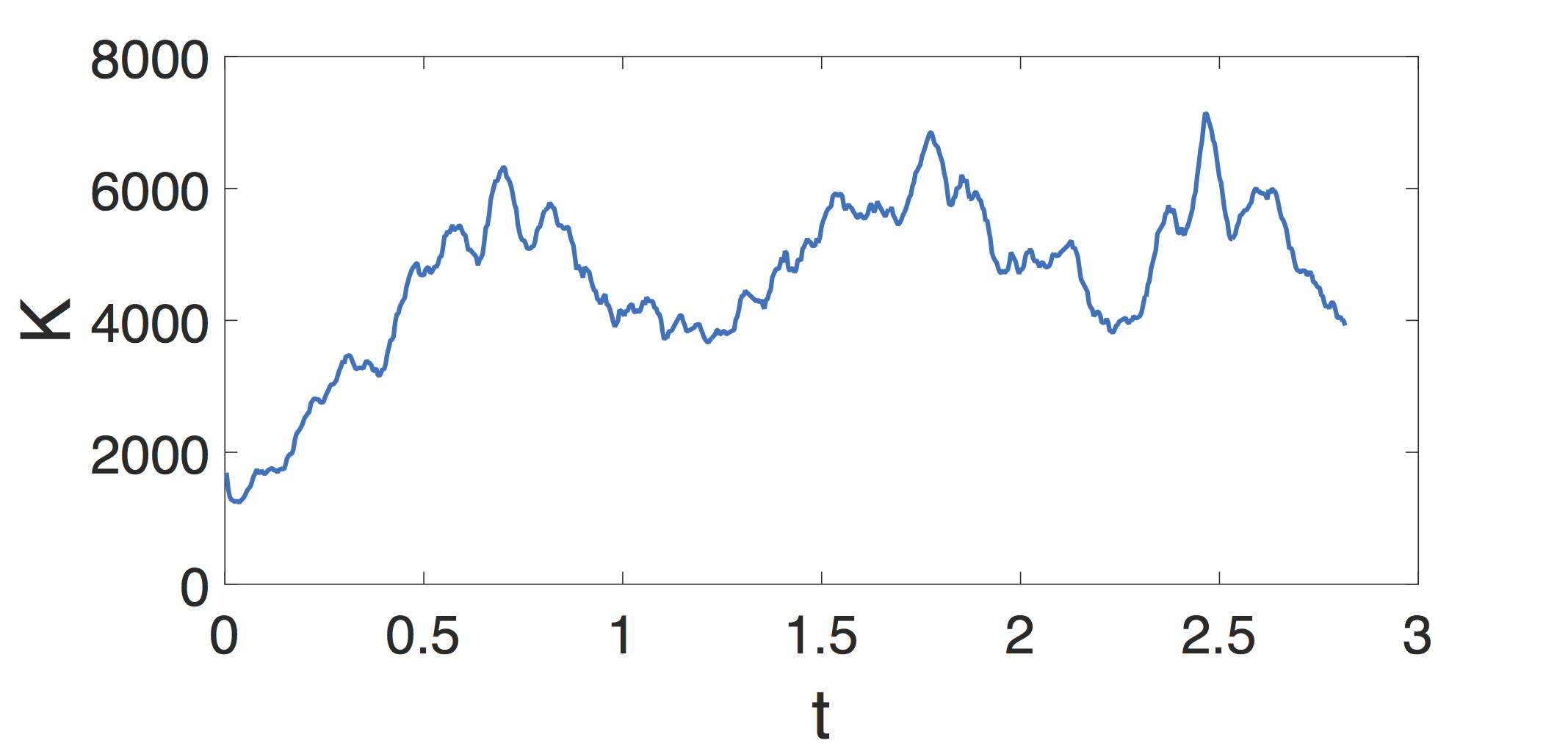}
\caption{Number of crossings, $k$, as a function of time $t$ (in $\rm s$). The average number of crossings for $t > 0.6$ is ${\bar k} = 4678.6$ }
\label{fig6}
\end{figure}

\newpage
\begin{figure}%
\centering
\includegraphics[width=0.9\linewidth]{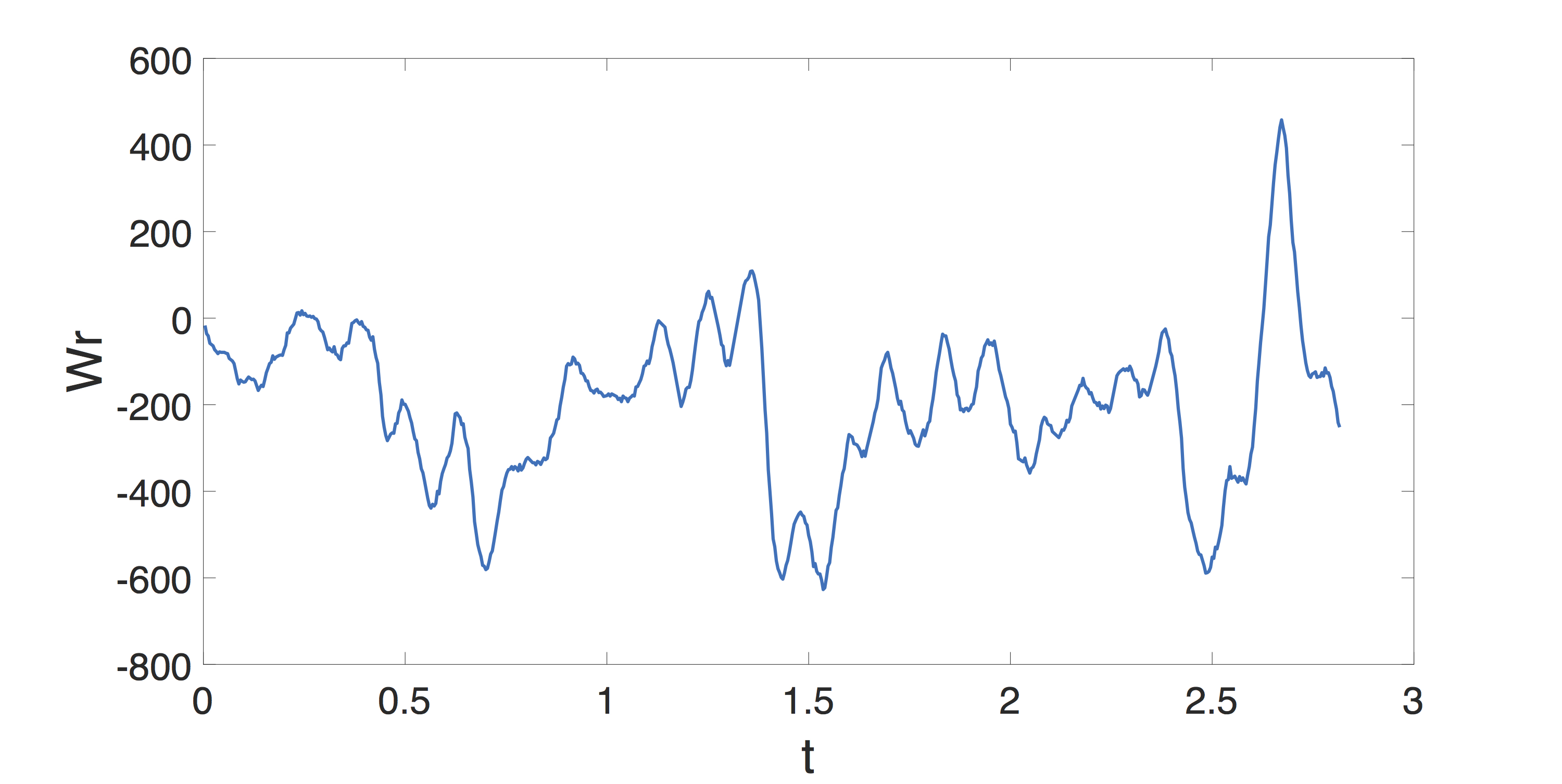}
\caption{Writhing number of the vortex tangle, $Wr$, as a function of 
time $t$  (in $\rm s$). The average writhing number for $t > 0.6$ is
%${\bar Wr} = -202.52$.}
${\bar Wr} = -205.3$.}
\label{fig7}
\end{figure}

\newpage
\begin{figure}%
\centering
\includegraphics[width=0.9\linewidth]{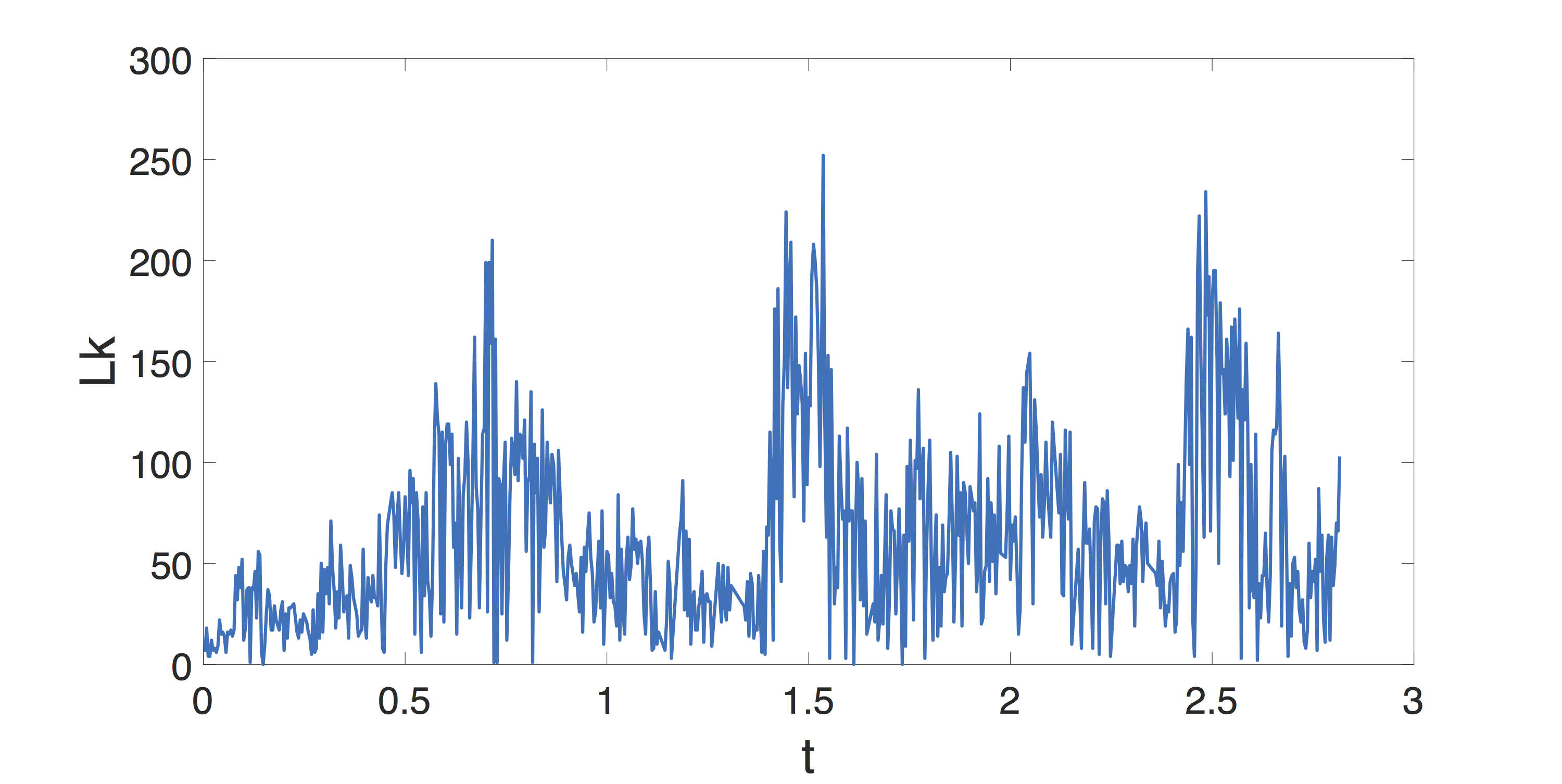}
\caption{Total linkage of the tangle, $Lk$, as a function of 
time $t$ (in $\rm s$). 
%The average linkage ${\bar Lk} = 62.54$.
The average linkage for $t > 0.6$ is ${\bar{Lk}} = 54.0$.
}
\label{fig8}
\end{figure}

\newpage
\begin{figure}%
\centering
\includegraphics[width=0.9\linewidth]{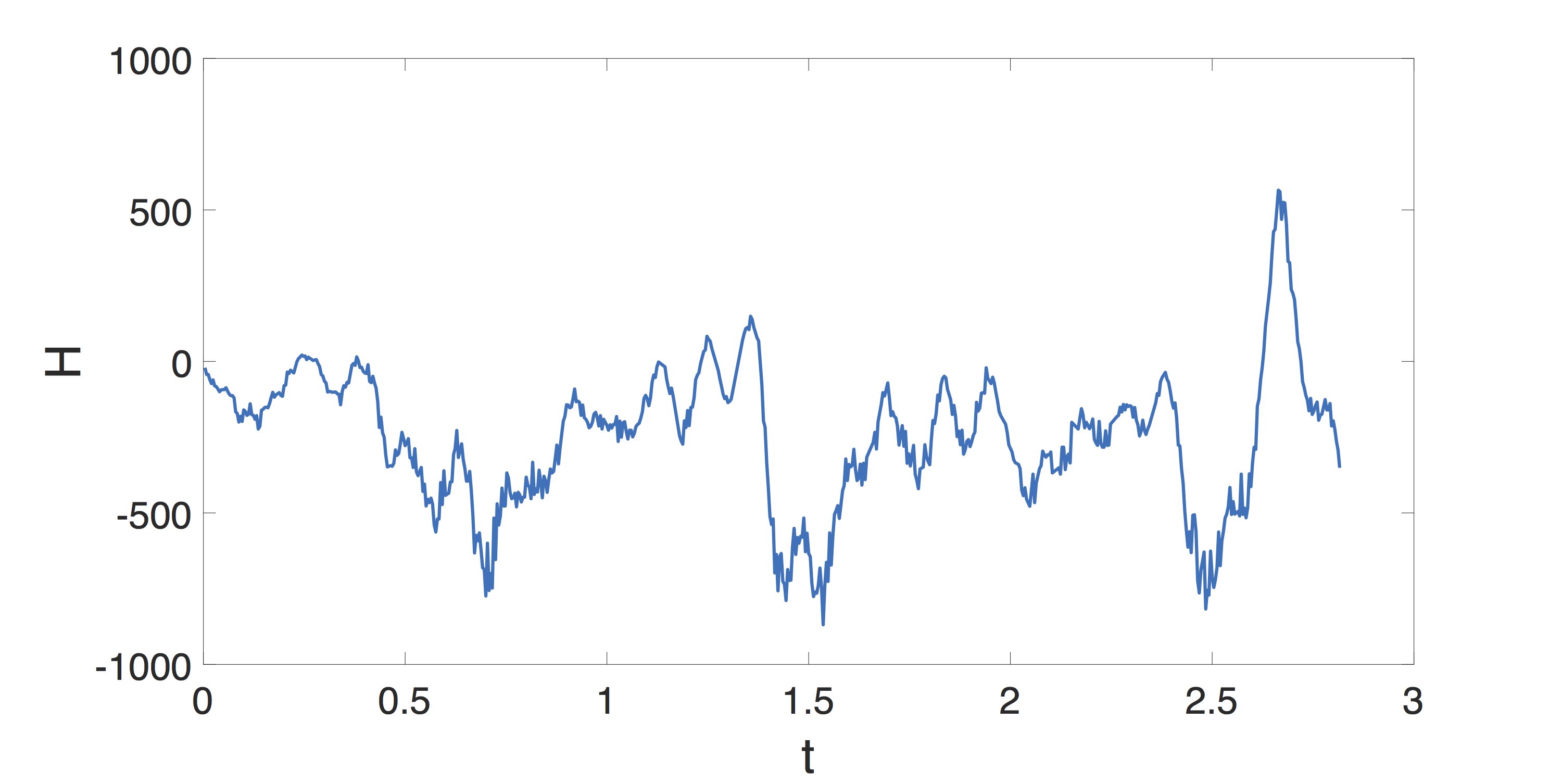}
\caption{Centreline helicity of the tangle, $H$, 
(in units of $\kappa^2$) as a function of time $t$ (in $\rm s$).
%The average centreline helicity is ${\bar H}= -96.39$
The average centreline helicity for $t > 0.6$ is ${\bar H}= -325.6$.
}
\label{fig9}
\end{figure}

%\newpage
%\begin{figure}[htp]
%\centering
%\includegraphics[width=0.9\linewidth]{figure10.eps}
%\caption{
%Probability density function (PDF) representing the
%distribution of values of $\vert Wr_j \vert$, 
%the magnitude of the writhing number of each loop $j$, at
%different times. 
%Different times are distinguished by different
%colours and symbols joined by a dashed line: 
%$t=4.0~\rm s$ (light red circles), 
%$t=4.04~\rm s$ (light red squares), 
%$t=4.044i~\rm s$ (light red triangles), 
%$t=4.048~\rm s$ (dark blue circles), 
%$t=4.052~\rm s$ (dark blue squares), 
%5$t=4.056~\rm s$ (dark blue triangles), 
%$t=4.06~\rm s$ (light blue circles), 
%$t=4.064~\rm s$ (light blue squares), 
%$t=4.068~\rm s$ (light blue triangles), 
%$t=4.072~\rm s$ (dark red circles), 
%$t=4.076~\rm s$ (dark red squares), 
%$t=4.08~\rm s$ (dark red triangles), 
%$t=4.084~\rm s$ (dark purple circles), 
%$t=4.088~\rm s$ (dark purple squares), 
%$t=4.092~\rm s$ (dark purple triangles), 
%$t=4.096~\rm s$ (light purple circles), 
%$t=4.1~\rm s$ (light purple squares), 
%$t=4.104~\rm s$ (light purple triangles).
%Most loops have $Wr_j=0$,
%but, at any time, there is always a small but finite probability
%of finding a loop with large value if writhe.
%}
%\label{fig9}
%\end{figure}

\newpage
\begin{figure}
\centering
\includegraphics[width=0.9\linewidth]{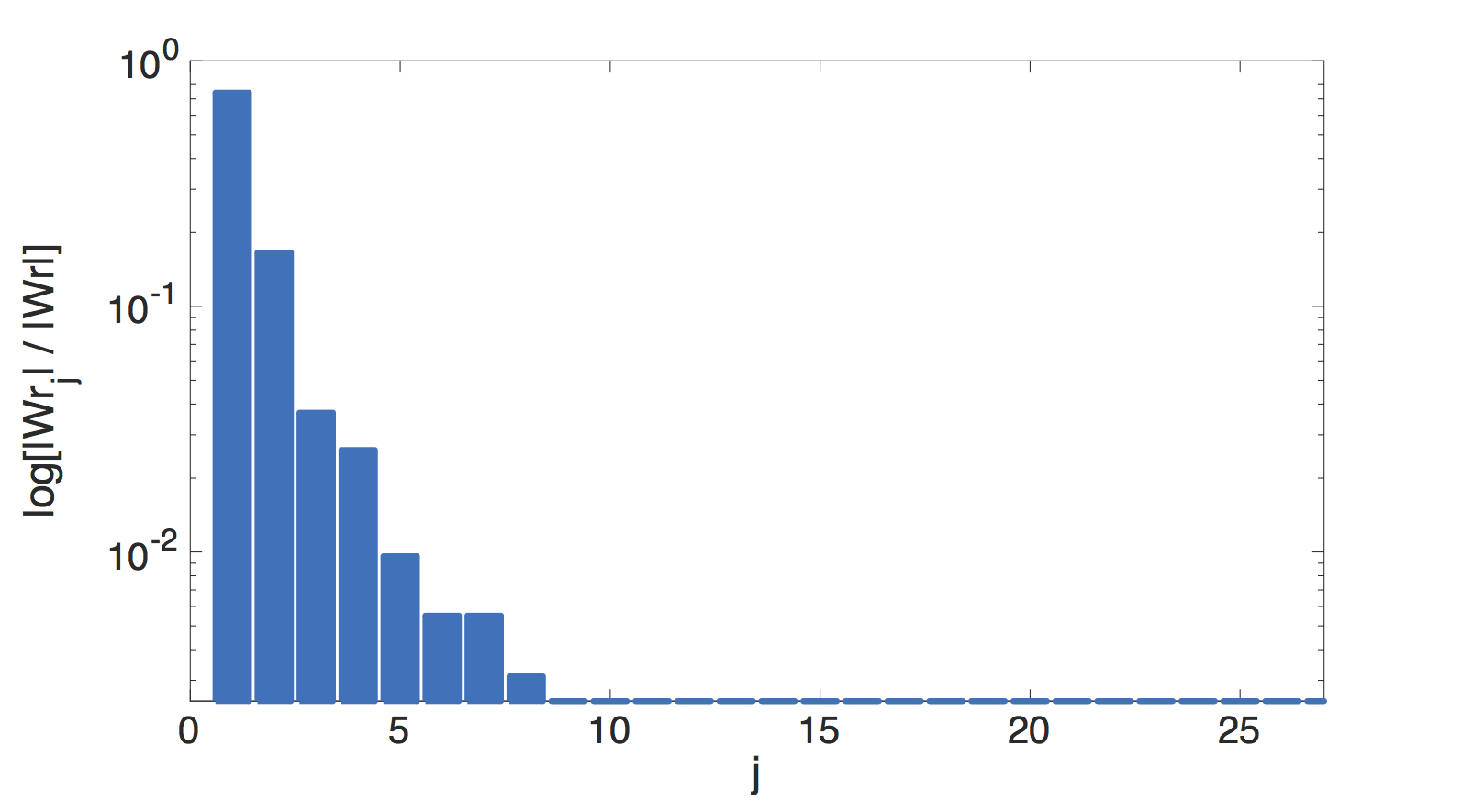}
\caption{
Relative distribution of writhe, $\log{(\vert Wr_j \vert)/\vert Wr\vert}$,
 obtained by averaging over 50 vortex
configurations at different times in the saturated regime. At each time,
the vortex loop $j=1$ has the largest writhe, the
loop $j=2$ has the second largest writhe, etc. It is apparent that the
total writhe of the vortex configuration is due to few large loops.}
\label{fig10}
\end{figure}

\newpage
\begin{figure}
\centering
\includegraphics[width=0.45\linewidth]{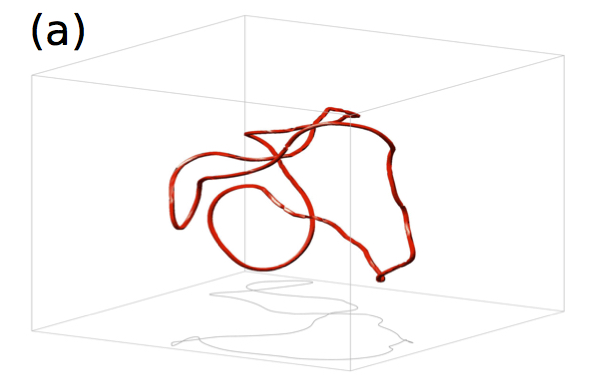}\qquad
\includegraphics[width=0.45\linewidth]{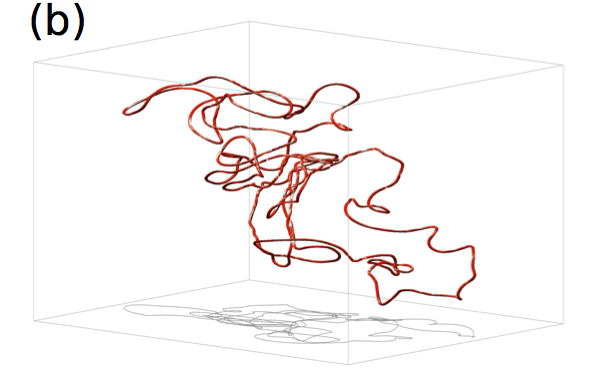}\\
\includegraphics[width=0.45\linewidth]{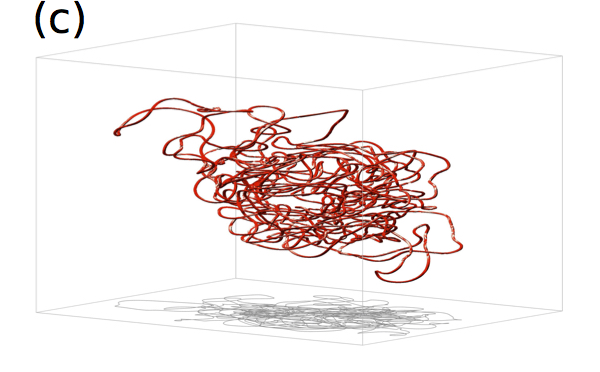}\qquad
\includegraphics[width=0.45\linewidth]{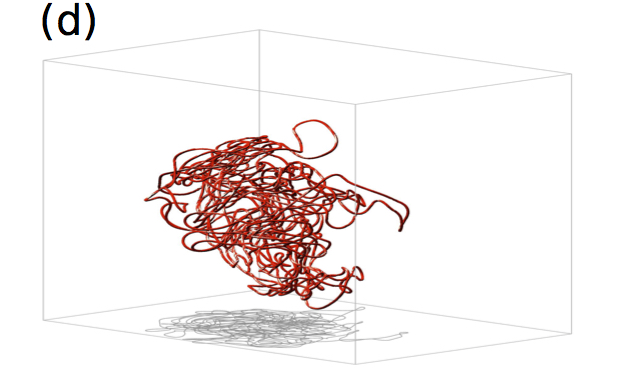}\\
\includegraphics[width=0.45\linewidth]{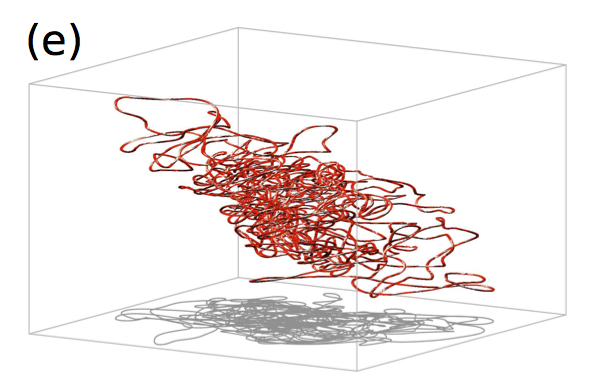}\qquad
\includegraphics[width=0.45\linewidth]{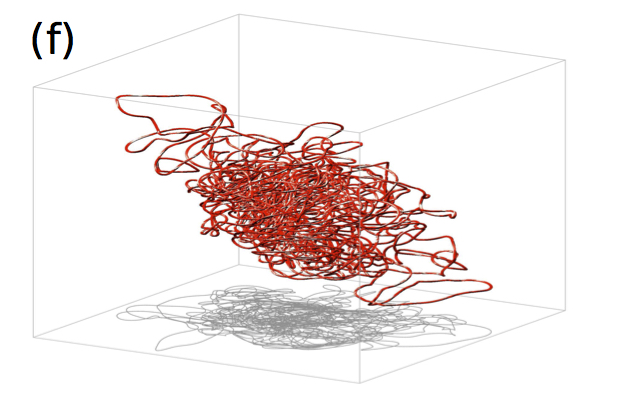}
\caption{
Examples of vortex loops with given writhing number $\vert Wr_j \vert$. (a): A vortex loop of $0$ writhe. (b): This vortex loop has $\vert Wr_j \vert =9$. (c): The writhe of this vortex loop is is $21$. (d): This loop has a slightly larger writhe of $\vert Wr_j \vert =92$. (e): This loop which is the largest at the time has writhe $124$. (f): This example is a vortex loop with one of the highest writhes in the whole system with $\vert Wr_j \vert =124$. 
}
\label{fig11}
\end{figure}

%\newpage
%\begin{figure}%
%\centering
%\includegraphics[width=0.9\linewidth]{figure12.eps}
%\caption{PDF (probability density function) representing
%the distribution of values of $A$, the order of the Alexander 
%polynomial of each vortex loop of the tangle at a given time
%represented by different symbols and colours joined by a dashed line:
%$t=0.7~\rm s$ (light red circles), 
%$t=0.908~\rm s$ (light red squares), 
%$t=1.1~\rm s$ (light red triangles), 
%$t=1.3~\rm s$ (dark blue circles), 
%$t=1.508~\rm s$ (dark blue squares), 
%$t=1.704~\rm s$ (dark blue triangles), 
%$t=1.9~\rm s$ (light blue circles), 
%$t=2.104~\rm s$ (light blue squares), 
%$t=2.3~\rm s$ (light blue triangles), 
%$t=2.504~\rm s$ (dark red circles), 
%$t=2.7~\rm s$ (dark red squares), 
%$t=2.9~\rm s$ (dark red triangles), 
%$t=3.1~\rm s$ (dark purple circles), 
%$t=3.312~\rm s$ (dark purple squares), 
%$t=3.5~\rm s$ (dark purple triangles), 
%$t=3.696~\rm s$ (light purple circles), 
%$t=3.896~\rm s$ (light purple squares), 
%$t=4.092~\rm s$ (light purple triangles).
%The peak at $A=0$ suggests that most vortex loops are unknots. Note however
%that at each time there is always a finite probability of finding a loop
%with Alexander polynomial of very high order.
%}
%\label{fig11}
%\end{figure}

\newpage
\begin{figure}%
\centering
\includegraphics[width=0.9\linewidth]{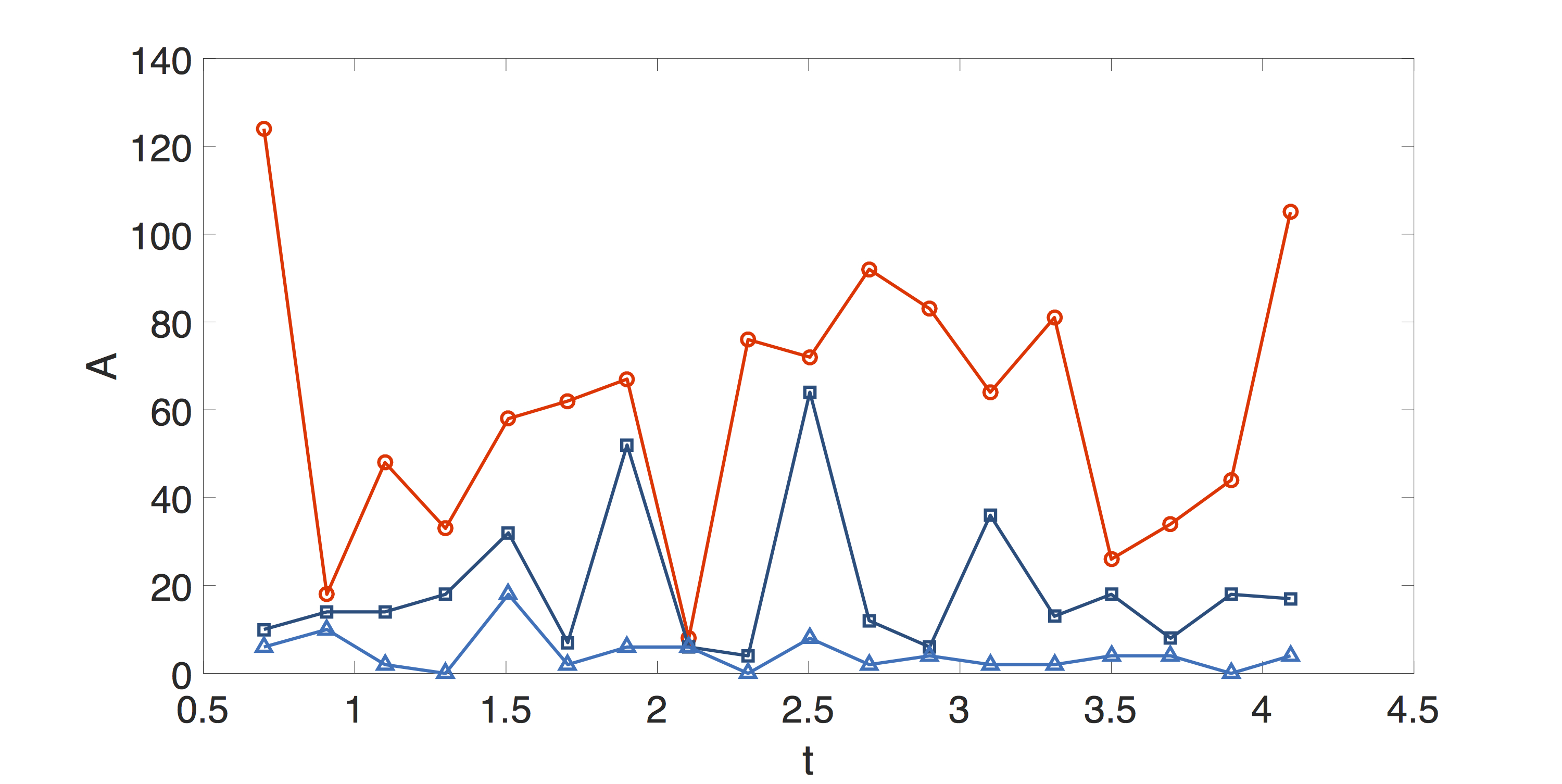}
\caption{The evolution of the 1st (red), 2nd (dark blue) and 3rd (light blue)
highest orders of Alexander polynomials with time $t$ ($\rm s$).}
\label{fig12}
\end{figure}

\newpage
\begin{figure}%
\centering
\includegraphics[width=0.43\linewidth]{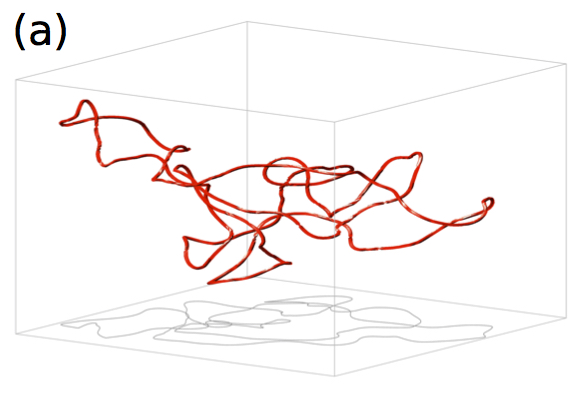}\qquad
\includegraphics[width=0.47\linewidth]{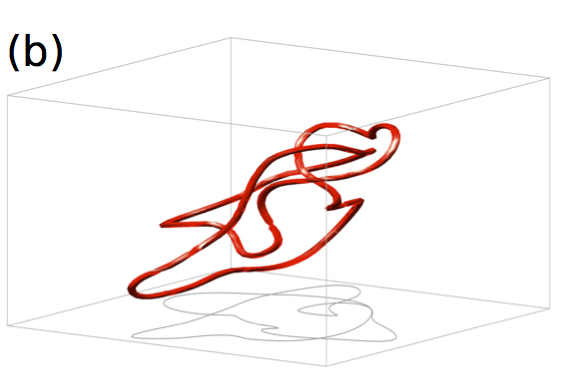}\\
\includegraphics[width=0.44\linewidth]{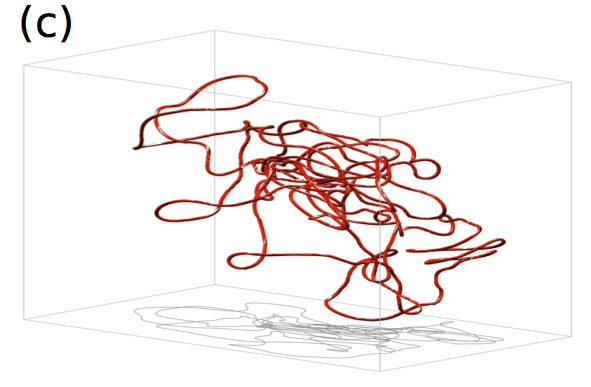}\qquad
\includegraphics[width=0.46\linewidth]{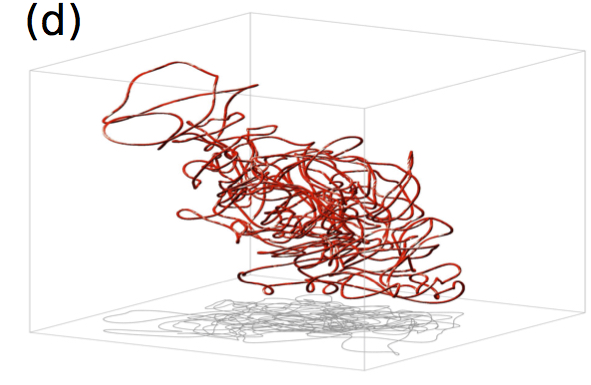}\\
\includegraphics[width=0.45\linewidth]{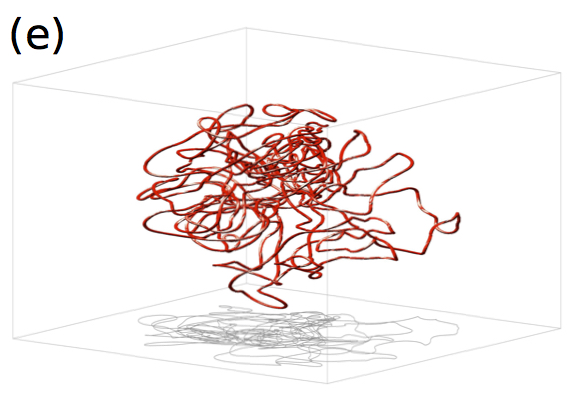}\qquad
\includegraphics[width=0.45\linewidth]{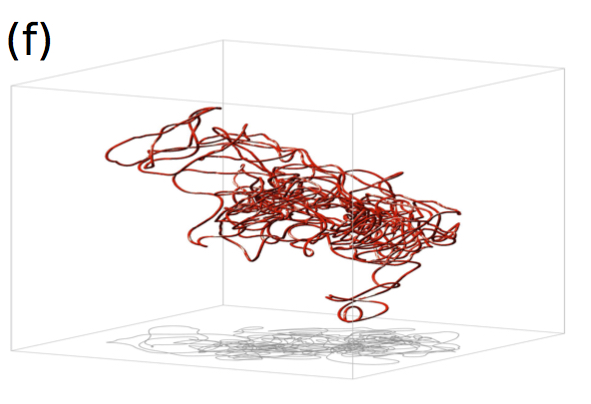}
\caption{Examples of vortex loops with given Alexander polynomial 
of order A taken from the numerical simulation. 
(a): A complicated-looking loop which has Alexander polynomial  
$\bigtriangleup (\tau)=1$, so it may be an unknot. 
(b): This vortex loop has Alexander polynomial  
$\bigtriangleup (\tau)=1-\tau+\tau^2$ which is the polynomial of 
the trefoil ($3_1$) knot. In fact this particular vortex loop can be 
easily manipulated into a trefoil by hand. 
(c): The order of the Alexander polynomial for this loop is 4 
as for the Solomon's Seal ($5_1$) knot. 
(d): This loop has Alexander polynomial of order 18. 
(e): The Alexander polynomial of this vortex loop is of order 36 
(it is not even the highest of the tangle at this particular time).
(f): This loop is the largest loop in the tangle at this particular
time and has Alexander polynomial of order 83.}
\label{fig13}
\end{figure}

\begin{figure}%
\centering
\includegraphics[width=0.9\linewidth]{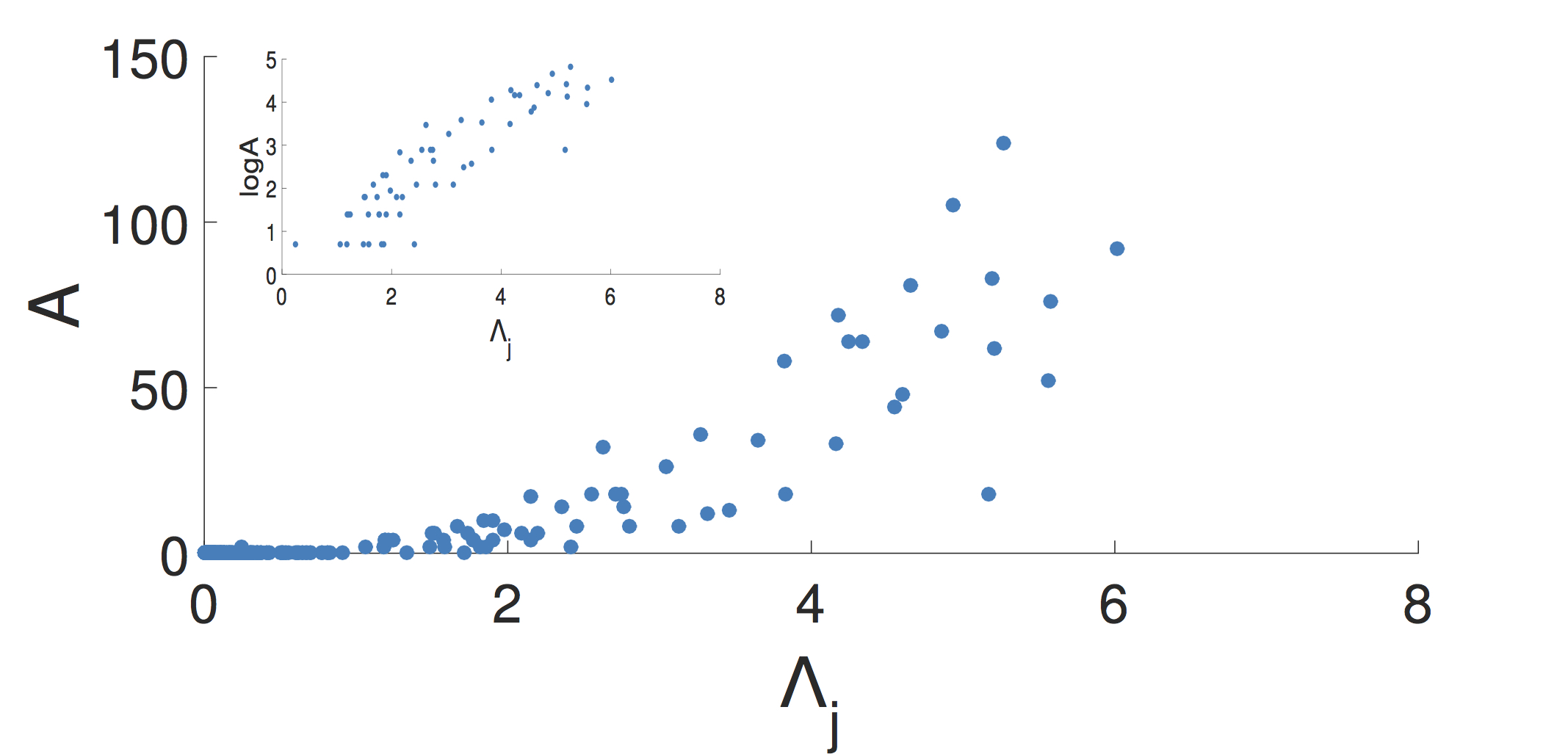}
\caption{The relationship between the length of a vortex loop, $\Lambda_j$
($\rm cm$), and the order of its Alexander polynomial, $A$
(data taken from the time evolution of a single realization).}
\label{fig14}
\end{figure}

%%%%%%%%%%%%%%%%%%%%%%%%%%%%%%%%%%%%%%%%%%%%%%%%%%%%%%%%%%%
\end{document}